\documentstyle[12pt]{article}
\def\dspace{\baselineskip=0.3 in}
\begin{document}
\dspace

\centerline{\bf Tachyon driven solution to Cosmic Coincidence Problem }

\vspace{1cm}

\centerline{\bf S.K.Srivastava,}
\centerline{\bf Department of Mathematics,}

\centerline{\bf North Eastern Hill University,}

\centerline{\bf Shillong - 793022}

\centerline{\bf ( INDIA )}

\centerline{\bf e - mail : srivastava@nehu.ac.in ;}

\centerline{\bf sushil@iucaa.ernet.in}

\vspace{2cm}

\centerline{\bf Abstract}

Here, non-minimally coupled tachyon to gravity is considered as a source of
$dark$ $energy$. It is demonstrated that with expansion of the universe,
$tachyon$ $dark$ $energy$ decays to $dark$ $matter$ providing a solution to
``cosmic coincidence problem''. Moreover, it is found that the universe
undergoes accelerated expansion simultaneously.  PACS nos. 98.80 Cq, 95.35.+d.

\vspace{2cm}

\centerline{\bf 1. Introduction}

\bigskip

Around four decades back, tachyons were proposed [1] and cosmology, driven by
these particles, were explored [2]. But these superluminal particles were
discarded for not being observed. At the turn of last century, tachyons are
re-awakened in the context of unstable D-branes in bosonic and superstring
theories. Due to concerted efforts by Sen [3], role of tachyons, in string
theories, got prominence among physicists. The idea is derived from the fact
that usual open string vacuum is unstable but with a stable vacuum also having
vanishing energy density. The unstable vacuum corresponds to $rolling$
$tachyon$ from the maximum of its potential to the minimum and stable vacuum
indicates presence of standard particle. Sen has argued that tachyonic state
is analogous to condensation of electric flux tubes of closed strings
described by Born - Infeld action. So, flat space Born - Infeld lagrangian was
suggested for tachyon condensates too [3, 4]. It was translated to curved
space framework, for tachyon scalars $\phi$ with potential $V(\phi)$, as $ -
V(\phi)\sqrt{ 1 -   g^{\mu\nu} \partial_{\mu} \phi \partial_{\nu} \phi }$
having minimal coupling with gravity [5].  Later on, Bagla $et$ $al.$ had
shown that this lagrangian could also be treated as generalization of a
relativistic particle lagrangian [6]. Recently, another tachyon model has been
proposed with lagrangian $ W(\phi)\sqrt{    g^{\mu\nu} \partial_{\mu} \phi
  \partial_{\nu} \phi - 1 }$  (with $W(\phi)$ is real). It is argued that tachyon scalars, described by this lagrangian, may be able to explore more physical situations than $quintessence$ [7]. It has been shown that as a homogeneous tachyon rolls down the hill of its potential to its minimum, when ${\dot \phi} = {d\phi}/{dt} \to 1,$ energy density  approaches a finite value and pressure tends to zero. In the cosmological framework, rolling of tachyon is associated with expansion of the universe [5]. These results prompted to conclude that when cosmic expansion is large, tachyon condensates behave like dust, showing it as good candidate for $cold$ $dark$ $matter$ (CDM), being pressureless non-baryonic fluid [8].

 Drastic changes are noticed on taking non-minimal coupling of $\phi$, given by the lagrangian [9]

$$ L_{\phi} =  \sqrt{- g} \Big[  - V (\phi) \sqrt{ 1 -
  g^{\mu\nu} \partial_{\mu} \phi \partial_{\nu} \phi  + \xi R \phi^2}\Big],
  \eqno(1.1)$$
where $R$ is the Ricci scalar, $\xi $ is the non-minimal coupling constant,
  $V(\phi)$ is the potential and $G = M_P^{-2}$ ( $M_P = 10^{19} {\rm GeV}$
  being the Planck mass) is the gravitational constant. Here $g$ is the
  determinant of metric tensor components $ g_{\mu\nu}(\mu, \nu = 0,1,2,3)$.

From this lagrangian, it is found that pressure is non-zero if $\xi \ne 0$
even when ${\dot \phi} \to 1.$

 Non-minimal coupling of tachyon with gravity was also proposed by Piao $et$ $al$ [10] in a different manner, where a function of $\phi$ is coupled to Einstein-Hilbert lagrangian as

$$ S = \int {d^4x} \sqrt{- g} \Big[ \frac{f(\phi) R}{16 \pi G} + V (\phi) \sqrt{ 1 + \alpha^{\prime}
  g^{\mu\nu} \partial_{\mu}\phi \partial_{\nu}\phi } \Big],
  \eqno(1.2)$$
where $\alpha^{\prime}$ gives string mass scale. Subject to the condition $
  1 - g^{\mu\nu} \partial_{\mu} \phi \partial_{\nu} \phi  >> \xi R \phi^2,$
  the lagrangian(1.1) looks like

$$ L_{\phi} \simeq  \sqrt{- g} \Big[  - V (\phi) \sqrt{ 1 -
  g^{\mu\nu} \partial_{\mu} \phi \partial_{\nu} \phi}  - \frac{1}{2} \frac{\xi
  V (\phi)\phi^2 R}{\sqrt{1 -
  g^{\mu\nu} \partial_{\mu} \phi \partial_{\nu} \phi}} ,
    \eqno(1.3)$$
which is similar to lagrangian of (1.2) with non-minimal coupling function
  $f(\phi)= - 8 \pi G \frac{\xi V (\phi)\phi^2 }{\sqrt{1 -
  g^{\mu\nu} \partial_{\mu} \phi \partial_{\nu} \phi}} $.

In what follows, investigations are made using the lagrangian (1.1). With this
lagrangian, tachyon condensates never approach to zero pressure. So, it can be
taken as viable candidate for $dark$ $energy$ (DE), not CDM. Apart from
tachyons, various DE models are available in the literature (i) a very small
cosmological constant [11], (ii) quintessqnce [12], (iii) k-essence [13],
(iv)Chaplygin gas [14], (v)interacting quintessence [15], (vi) non-minimally
coupled quintessence [16] and others, violating `` strong energy
condition''. Recently, some other models were also proposed violating `` weak
energy condition'' also [17] and showing finite time future
singularities. Barrow [18] and Lake [19] have demonstrated that  violation of
``dominant energy condition'' leads to `` sudden future singularity''. A
generalization of Barrow's model is suggested in ref.[20]  . A review on DE can be seen in articles [21, 22].

The astrophysical observation that high red-shift supernovae are fainter than
expected, leads to cosmic acceleration [23-25]. This observation is further
confirmed by WMAP (Wilkinson  Microwave Anisotropy Probe) [26-28]. These
experiments show that  content of the present universe is comprised of around
77\% dark energy and 23\% CDM. In addition to accelerated expansion of the current universe, these data pose another important question `` How are DE and CDM densities of the same order at present ?'' This question constitutes the ``cosmic coincidence problem''(CCP)[29]. As mentioned above, experimental data also suggest that DE dominates over CDM today. So, in other words, CCP is also coined as `` How does DE dominate the present universe ?''

The accelerated expansion is caused by DE, driven by different exotic matter mentioned above. Moreover, for ``coincidence problem'' too, role of DE is crucial. So, it is important to understand its nature. In the recent past, it was repeatedly suggested that DE could be derived by self-interacting scalar fields, behaving as a perfect fluid with the equation of state parameter w$< - 1/3.$ It was proposed that, in the early universe, these scalars contributed DE density lower than that of matter including radiation. With the expansion of the universe, matter density rolled down. As a result, DE density bacame comparable to matter density at late times [12]. The usual strategy, in these types of work, is the use of a suitable potential to yield the required result. Later on, Padmanabhan [30] demonstrated that it is straightforward to derive such potentials. In ref.[31],it is shown that a mixture of matter and $quintessence$, gravitationally interacting with each other, is unable to derive speeded-up expansion and a solution to CCP simultaneously unless matter fluid is dissipative enough. These authors have obtained attractor solution for $r = \rho_{\phi}/\rho_m $(with $ \rho_{\phi}(\rho_m) $ being DE (matter)density). Here, it is shown that $r < 1$ remains stable [15,32].

In ref.[33], author has proposed a different mechanism to overcome CCP alongwith accelerated expansion. Contrary to assumption in refs.[12]. This prescription suggests that, in the early universe, DE density used to be high, but dynamical. Moreover, it is shown that DE density dominates the early universe , causing accelerated expansion from the beginning of the universe . This situation is like inflationary models with the difference that inflationary models exhibit accelerated expansion for a short period, whereas in the model of ref.[33] expansion is slowly speeded-up from the epoch of creation of the universe upto late times. DE density falls down with the growth of scale factor. DE, so lost, causes creation of $dark$ $matter$ (DM). Accordingly, in this model,there is no DM in the beginning, rather it is produced due to decay of DE. So, $r$ becomes a dynamical parameter and grows with the time keeping itself as  $0 \le r < 1.$ The phenamenon of decay of DE and creation of DM is given by coupled equations. A similar procedure has been adopted by Mota and Bruck [34] for the condensation of DE in overdense regions of matter, though it does not condensate in normal circumstances. 
  
The same mechanism [33] is adopted in the present paper for tachyon
condensates, in the late universe, as in [9],where tachyon DE decays
to CDM. In this paper, self interacting inverse cubic potential had been considered. Here, investigations for tachyons are carried out taking self interacting inverse quartic potential.

The paper is organized as follows. Using action (1.1), basic equations are derived in section 2. Section 3 contains investigations employing inverse quartic potential with a modification in coupled equations for DE and DM. Section 4 deals with inverse exponential potential. Remarks on the results, obtained in sections 3 and 4, are given in section 5. Natural units $\hbar =c = 1,$ are used with GeV as a fundamental unit.
   
\bigskip

\centerline{\bf 2. Basic equations}

\bigskip

 Einstein's field equations are given as

$$ R_{\mu\nu} - \frac{1}{2}g_{\mu\nu} R = - 8 \pi G [ T_{\mu\nu (\phi)} + T_{\mu\nu (m)}]
\eqno(2.1)$$ 
with energy-momentum tensor components of tachyon and matter as
$$ T_{\mu\nu (\phi)} = (\rho_{(\phi)} + p_{(\phi)})u_{\mu} u_{\nu} - p_{(\phi)}
  g_{\mu\nu}  \eqno(2.2a)$$
and
$$ T_{\mu\nu (m)} = (\rho_{(m)} + p_{(m)})u_{\mu} u_{\nu} - p_{(m)}
  g_{\mu\nu}  \eqno(2.2b)$$
respectively, where $u^{\mu} = (1,0,0,0). T^{\mu}_{\mu (\phi)} =
  (\rho_{(\phi)}, -  p_{(\phi)}, - p_{(\phi)}, - p_{(\phi)})$ are obtained
  from the lagrangian (1.1) with

\begin{eqnarray*}
 T_{\mu\nu (\phi)} &=& - V (\phi) [ 1 -
   \bigtriangledown^{\rho} \phi \bigtriangledown_{\rho} \phi  + \xi R
   \phi^2]^{-1/2} \times \Big[ - \bigtriangledown_{\mu} \phi
   \bigtriangledown_{\nu} \phi + \xi R_{\mu\nu} \phi^2 \\ &&  + \xi (
   \bigtriangledown_{\mu} \bigtriangledown_{\nu} - g_{\mu\nu} {\Box} ) \phi^2
   - g_{\mu\nu} ( 1 -
   \bigtriangledown^{\rho} \phi \bigtriangledown_{\rho} \phi  + \xi R
   \phi^2) \Big].
\end{eqnarray*}
$$ \eqno(2.2c)$$
Here $\bigtriangledown_{\mu}$ stands for covariant derivative and $R_{\mu\nu}$
are Ricci tensor components.

Field equations for $\phi$ are obtained as 

$${\Box} \phi + \frac{2(\bigtriangledown^{\mu} \phi) (\bigtriangledown_{\rho}
  \phi) (\bigtriangledown^{\rho} \bigtriangledown_{\mu} \phi) - 2 \xi R \phi \bigtriangledown^{\rho} \phi \bigtriangledown_{\rho} \phi - \xi \phi^2 g^{\mu\nu}\bigtriangledown_{\mu} R \bigtriangledown_{\nu} \phi}{2 (1 -
   \bigtriangledown^{\rho} \phi \bigtriangledown_{\rho} \phi  + \xi R
   \phi^2)} $$
$$ + \xi R \phi + \frac{V^{\prime}}{V} ( 1 + \xi R \phi^2 ) = 0, \eqno(2.3)$$ 
from the lagrangian (1.1). Here $V^{\prime} (\phi) = \frac{d}{dx} V(\phi)$ and 
$${\Box} = \bigtriangledown^{\rho} \bigtriangledown_{\rho} = \frac{1}{\sqrt{-g}}\frac{\partial}{\partial x^{\mu}} ( \sqrt{-g} g^{\mu\nu} \frac{\partial}{\partial x^{\nu}}).$$

 According to cosmological observations [23 - 28], currently we live in a spatially flat and speeding - up universe, such that
${\ddot a}/ a > 0$ for the scale factor $a(t)$, given by the distance function

$$ dS^2 = dt^2 - a^2(t) [ dx^2 + dy^2 + dz^2 ]. \eqno(2.4)$$
It represents a homogeneous model of the universe, hence 

$$ \phi (x, t) = \phi (t) .\eqno(2.5)$$

Connecting eqs.(2.3) and (2.5a)

$${\ddot \phi} + 3 H {\dot \phi} + \frac{2 {\ddot \phi}{\dot \phi}^2 - 2 \xi R
  \phi {\dot \phi}^2 - \xi \phi^2 {\dot R}{\dot \phi}}{2 ( 1 - {\dot \phi}^2 +
  \xi R \phi^2 ) } +  \xi R \phi + \frac{V^{\prime}}{V} ( 1 + \xi R \phi^2 ) =
  0 , \eqno(2.6)$$
where $H = {\dot a}/a .$

Energy density $\rho_{(\phi)}$ and pressure $p_{(\phi)}$ are obtained from
eq.(2.2c) as 
\begin{eqnarray*}
\rho_{\phi} &=&  V(\phi) \frac{\Big[ 1 - \xi \Big(R^0_0 - R \Big)\phi^2 + 6
  \xi H \phi {\dot \phi} \Big]}{\sqrt{1 - {\dot \phi}^2 + \xi \phi^2 R}}
  \\&=& V(\phi) \frac{[ 1 +  6
  \xi H \phi {\dot \phi}  + 3 \xi ({\dot H} + 3 H^2) \phi^2
  }{\sqrt{1 - {\dot \phi}^2 + 6 \xi \phi^2 ({\dot H} + 2 H^2)}}
\end{eqnarray*}
$$ \eqno(2.7a)$$
and isotropic pressure $p_{\phi}$ as
\begin{eqnarray*}
p_{\phi} &=& -  V(\phi) \frac{\Big[ 1 - {\dot \phi}^2  + \xi (2 \phi {\ddot
    \phi} + 2 {\dot \phi}^2 + 6
  \xi H \phi {\dot \phi}) + \xi \Big( R^1_1
  - R \Big)\phi^2 \Big]}{\sqrt{1 - {\dot \phi}^2 \xi \phi^2 R}} \\&=&  -  V(\phi) \frac{\Big[ 1 - {\dot \phi}^2  + \xi (2 \phi {\ddot
    \phi} + 2 {\dot \phi}^2 + 6
  \xi H \phi {\dot \phi}) + \xi( 5 {\dot H} + 9 H^2 )\phi^2 \Big]}{\sqrt{1 - {\dot \phi}^2 \xi \phi^2 R}}
\end{eqnarray*}
$$ \eqno(2.7b)$$

From eqs.(2.1), it is obtained that

$$\frac{R^1_1 - \frac{1}{2}R}{R^0_0 - \frac{1}{2}R} \simeq \frac{- p_{\phi}}{\rho_{\phi}} \eqno(2.8)$$
taking dominance of tachyon dark energy over matter. Here $\rho_{\phi}$ and
$p_{\phi}$ are given by eqs.(2.7).

Bianchi identities $(T^{\mu\nu}_{ (\phi)} + T^{\mu\nu}_m )_{:\nu} = 0$
yield coupled equations
$${\dot \rho_{\phi}} + 3 H (\rho_{\phi} + p_{\phi}) = - Q(t) \eqno(2.9a)$$
and
$${\dot \rho_m} + 3 H \rho_m =  Q(t),  \eqno(2.9b)$$
where $Q(t)$ is loss (gain) term for DE (CDM). Here $p_m = 0$
for CDM. Physically, these equations show decay of DE to CDM.

\bigskip

\centerline{\bf 3. Inverse self-interacting quartic potential for tachyon}

\bigskip

The inverse quartic potential for $\phi$ is taken as

$$ V(\phi) = \lambda \phi^{- 4}, \eqno(3.1)$$
where $\lambda$ is a dimensionless coupling constant.

\noindent \underline{Case (a) $\xi \ne 0$}

Using the potential, given by eq.(3.1), in eqs.(2.6), (2.7) and (2.8), it is
obtained as
$${\ddot \phi} + 3 H {\dot \phi} + \frac{2 {\ddot \phi}{\dot \phi}^2 - 2 \xi R
  \phi {\dot \phi}^2 - \xi \phi^2 {\dot R}{\dot \phi}}{2 ( 1 - {\dot \phi}^2 +
  \xi R \phi^2 ) } - 3 \xi R \phi -  \frac{4}{V}  =   0  \eqno(3.2a)$$
and

$$\frac{ 2 {\dot H} + 3 H^2}{3 H^2} \simeq \frac{\Big[ 1 - {\dot \phi}^2  + \xi (2 \phi {\ddot
    \phi} + 2 {\dot \phi}^2 + 6
  \xi H \phi {\dot \phi}) + \xi ( 5 {\dot H} + 9 H^2 )\phi^2 \Big]}{[ 1 +  6
  \xi H \phi {\dot \phi}  + 3 \xi ({\dot H} + 3 H^2) \phi^2 
  } = - {\rm w}_{\phi} \eqno(3.2b)$$
for the geometry given by eq.(2.4).

According to the tachyon lagrangian (1.1),  $\phi$ has mass dimension equal to $-1$ like time
$t$ (in natural units) . So, on the basis of dimensional considerations, it is
reasonable to take

$$  \phi (t) = A t .\eqno(3.3)$$
with $A$ being dimensionless constant.

Now the scale factor is assumed to have the form

$$ a(t) = a_i \Big(t/t_i \Big)^q, \eqno(3.4)$$
where $q$ is a real number and $t_i$ is supposed to be the time when DE begins
to decay to CDM. Also $a_i = a(t_i).$

Connecting eqs.(3.2)-(3.4), it is obtained that

$$\frac{3 q - 2}{3 q} \simeq \frac{1 - A^2 + \xi A^2 (2 + q + 9 q^2)}{1 + 3
  \xi A^2 (q + 3 q^2)} = - {\rm w}_{\phi} \eqno(3.5a)$$
and
$$36 \xi q^2 A^2 = 3 q A^2 ( 1 + 6 \xi ) - 4 . \eqno(3.5b)$$

Elimination of $A^2$, from eqs.(3.5),yields
$$ q = \frac{2}{3 (1 + {\rm w})_{\phi}} = \frac{14 \xi - 1}{4 \xi} . \eqno(3.6a,b)$$ 

Subject to the condition $ 0 < A^2 < [1 - 6 \xi (-q + 2 q^2]^{-1}$ to get
$\rho_{\phi}$ and $p_{\phi}$, eqs.(3.5) and (3.6) yield a set of solutions
$$ \xi = - 12.5 , \quad q = 3.52, \quad {\rm w})_{\phi} = - 0.81$$
and
$$ A^2 = \frac{100}{119856}. \eqno(3.7a,b,c,d)$$

From eqs.(2.7a), (2.9a), (3.5a) and (3.6), it is obtained that

$$ Q(t) = 2 {\tilde A} t^{-5}, \eqno(3.8a)$$
where
$${\tilde A} = \frac{\lambda}{A^4} [ 1 + 3 \xi A^2 (q + 3 q^2)]. \eqno(3.8b)$$

Using eqs.(3.4) and (3.7) , eq.(2.9b) is integrated to

$$ \rho_m = \frac{2 {\tilde A} t^{-4}}{(3 q - 4)} \Big[1 - \Big(t/t_i
\Big)^{(3q-4)} \Big]. \eqno(3.9)$$
for $\rho_m (t_i) = 0.$ From eq.(2.7a),(3.3) and (3.4)
$$\rho_{\phi} = {\tilde A} t^{-4} \eqno(3.10)$$
with ${\tilde A}$ given by eq.(3.7b). So,

$$ r(t) = \rho_m/\rho_{\phi} = 0.304 \Big[ 1 - \Big(t_i/t \Big)^{6.56} \Big]. \eqno(3.11)$$

Using current observational data for the universe, DE density $\rho_{\phi(0)}=
0.77 \rho_{\rm cr,0},$ CDM density $\rho_{m(0)}= 0.23 \rho_{\rm cr,0}$ with
$\rho_{\rm cr,0}= 3 H_0^2/8 \pi G, H_0 = h/t_0 (h = 0.72 \pm 0.04)$ and the
present age $t_0 \simeq 13.7$ Gyr, in eq.(3.10), it is obtained that

$$ t_i = 0.539 t_0. \eqno(3.12)$$

Eqs.(3.10) - (3.12) yield

$$ r(t) = 0.304 \Big[ 1 - 0.017 \Big(t_0/t \Big)^{6.56} \Big]. \eqno(3.13)$$
showing that $0 < r < 1$ for $0.539 t_0 < t$. Moreover as $q$ is greater than
1, eq.(3.4) shows that universe is accelerated. Thus it provides a solution to
CCP in the speeded-up universe.

\noindent \underline{Case (b) $\xi = 0$}

In this case, eqs.(3.2) look like

$$\frac{\ddot \phi}{ 1 - {\dot \phi}^2} + 3 H {\dot \phi} -  \frac{4}{V}  =
  0  \eqno(3.14a)$$
and

$$\frac{ 2 {\dot H} + 3 H^2}{3 H^2} \simeq  1 - {\dot \phi}^2  = - {\rm w}_{\phi} \eqno(3.14b)$$

Moreover eqs.(2.7) reduce to

$$\rho_{\phi} =  \frac{V(\phi)}{\sqrt{1 - {\dot \phi}^2}} \eqno(3.15a)$$
and
$$p_{\phi} = -  V(\phi) {\sqrt{1 - {\dot \phi}^2}} \eqno(3.15b)$$
 
Eqs.(3.15) show that ${\dot \phi}^2 < 1$ for real $\rho_{\phi}$ and
  $p_{\phi}$. Moreover, taking $ {\ddot \phi} \ll 3 H {\dot \phi}$ in
  eq.(3.14a), eqs.(3.14) are integrated to

$$ \phi = \phi_i \Big(t/t_i \Big)^{4/3 \sqrt{\lambda}} \eqno(3.16a)$$
and
$$ H(t) = \frac{3 \lambda}{8 \phi_i^2} t_i^{8/3 \sqrt{\lambda}} \Big[ - 1 +
\frac{8}{3 \sqrt{\lambda}} \Big] t^{(1 - 8/3 \sqrt{\lambda})} \eqno(3.16b)$$
with 
$$ \sqrt{\lambda} = - \frac{4}{3} - \frac{1}{3} \sqrt{10} . \eqno(3.16c)$$

As $\rho_{\phi} \simeq V(\phi),$ for ths case, $Q(t)$ is calculated to be 

$$ Q(t) = 2 \lambda \phi_i^{-4} t_i^{16/3 \sqrt{\lambda}} t^{(-1 - 16/3 \sqrt{\lambda})} \eqno(3.17)$$
using eq.(2.9a). Employing eqs.(3.16), eq.(2.9b) is integrated to 

$$\rho_m  \approx \frac{128 B}{9} \Big[(4 + \sqrt{10})/ (12 +
    \sqrt{10}) \Big]
    t^{- \sqrt{10}/(4 + \sqrt{10})} \Big[\Big(t/t_i)^{ \sqrt{10}/(4 + \sqrt{10})}$$
    $$\times exp\{B(12 + \sqrt{10}/16 + \sqrt{10})[ t^{(16 +
    \sqrt{10})/(4 + \sqrt{10})} - t_i^{(16 + \sqrt{10})/(4 + \sqrt{10})}\} - 1
    \Big], \eqno(3.18)$$
where $B = \frac{3 \lambda}{8 \phi_i^2} t_i^{8/3
    \sqrt{\lambda}}$. Now
$$\rho_m/\rho_{\phi} \simeq \rho_m/V(\phi) \approx \frac{16 \phi_i^2}{3}
t_i^{8/4 + \sqrt{10}} t^{- 16 - \sqrt{10}/(4 + \sqrt{10})} \Big[\Big(t/t_i)^{ \sqrt{10}/(4 + \sqrt{10})}$$
    $$\times exp\{B(12 + \sqrt{10}/16 + \sqrt{10})[ t^{(16 +
    \sqrt{10})/(4 + \sqrt{10})} - t_i^{(16 + \sqrt{10})/(4 + \sqrt{10})}\} - 1
    \Big]. \eqno(3.19)$$

As $t_0 > t_i$, eq.(3.19) shows $\rho_m/\rho_{\phi} > 1$. This result is $not$
consistent with current observations. It means that, in accelerated universe,
CCP can not be solved for $\xi = 0$ through decay of tachyon dark energy to CDM.

\bigskip

\centerline{\bf Conclusion}
\smallskip

In the above investigations, it is found that non-minimally coupled tachyon
with gravity contributes DE to the universe. Dynamics of tachyons and its
back-reaction to the universe are explored subject to  self-interacting
inverse quartic potential. Here, investigations for decay of DE to CDM are
made for cases of minimally coupled tachyon as well as non-minimally coupled
tachyon to gravity.  In the case of non-minimal coupling, 
present ratio of CDM and tachyon energy density is obtained to be $\sim
0.3$ in the accelerated universe without taking any dissipative term, if decay
of DE to CDM begins at $t_i = 0.539 t_0.$ But, in the case of minimal
coupling, CDM density is found more than tachyon energy denmsity contradicting
current observations. It means that, in the case of minimal coupling ($\xi
=0$) decay of tachyon dark energy to CDM is not possible. This result is
parallel to the result of ref.[15], where it is obtained that CCP can not be
solved in the accelerated universe unless dissipative term is used for CDM.

\bigskip

\centerline{\bf References}
\smallskip

\noindent [1] O.M.P.Bilaniuk, V.K.Deshpande and E.C.G.Sudarshan, Am. J. Phys. {\bf 30}, 718 (1962); O.M.P.Bilaniuk, V.K.Deshpande and E.C.G.Sudarshan, Physics Today {\bf 22}, 43 (1969); O.M.P.Bilaniuk and E.C.G.Sudarshan, Nature {\bf 223}, 386 (1969); G. Feinberg, Phys. Rev. {\bf 98}, 1089 (1967); E. Recami and R. Mignani, Riv. Nuovo Cimento, {\bf 4}, 209 (1974); A.F.Antippa and A. E. Everett, Phys. Rev. D {\bf 4}, 2098(1971); A.F.Antippa, Phys. Rev. D{\bf 11}, 724(1975); A. E. Everett, Phys. Rev. D, {\bf 13}, 785(1976); $idid$ {\bf 13}, 795(1976); K. H. Mariwalla, Am. J. Phys., {\bf 37}, 1281 (1969); L. Parker, Pys. Rev. {\bf 188}, 2287 (1969).
\bigskip

\noindent [2] J. C. Foster and J. R. Ray, J. Math. Phys. {\bf 13}, 979 (1972); J. R. Ray, Lett. Nuovo Cimento {\bf 12}, 249 (1975); S. K. Srivastava and M.P.Pathak, J. Math. Phys. {\bf 18}, 483 (1977); $ibid$ {\bf 18}, 2092 (1977); $ibid$ {\bf 19}, 2000 (1978); $ibid$ {\bf 23}, 1981 (1982); $ibid$ {\bf 24}, 966 (1983); $ibid$ {\bf 24}, 1311 (1983); $ibid$ {\bf 24}, 1317 (1983); $ibid$ {\bf 25}, 693 (1984).
\bigskip

\noindent [3] A. Sen, hep-th/9904207 ;J. High Energy Phys. {\bf 04}, 048 (2002) [hep-th/0203211];J. High Energy Phys. {\bf 07}, 065 (2002)[hep-th/0203265].; Mod. Phys. Lett. A {\bf 17}, 1799 (2002) and references therein.

\bigskip

\noindent [4] A. Sen, Mod. Phys. Lett. A {\bf 17}, 1799 (2002)[hep-th/0204143] and references therein.

\bigskip
\noindent [5] G. W. Gibbons, Phys. Lett. B {\bf 537}, 1 (2002)[hep-th/0204098]

\bigskip
\noindent [6] J. S. Bagla, H. K. Jassal and T. Padmanabhan, Phys. Rev. D {\bf 67}, 063504 (2003).

\bigskip

\noindent [7] V. Gorini, A. Kamenshchik, U. Moschella and V. Pasquir, Phys. Rev. D {\bf 69}, 123512 (2004); hep-th/0311111 ;  L. P. Chimento, Phys. Rev. D {\bf 69}, 123517(2003).

\bigskip

\noindent [8]  T. Padmanabhan, Phys. Rev. D {\bf 66}, 021301 (2002); T. Padmanabhan and T. R. Choudhury, Phys. Rev. D {\bf 66}, 081301 (2002); M. Fairbairn and M. H. G. Tytgat, Phys. Lett. B {\bf 546}, 1 (2002); A. Feinstein, D {\bf 66}, 063511 (2002); D. Choudhury, D. Ghosal, D.P.Jatkar and S. Panda, Phys. Lett. B {\bf 544}, 231 (2002); A. Frolov, L. Kofman and A. A. Starobinsky, Phys. Lett. B {\bf 545}, 8 (2002); L. Kofman and A. Linde, J. High Energy Phys. {\bf 07}, 065 (2002); G. Shiu and I. Wasserman,  Phys. Lett. B {\bf 541}, 6 (2002); L. R. W. Abramo and F. Finelli, Phys. Lett. B, {\bf 575}, 165 (2003). 
\bigskip

\noindent [9] S. K. Srivastava, gr-qc/0409074.
\bigskip

\noindent [10] Yun-Song Piao $et$ $al$., Phys. Lett. {\bf B 570}, 1 (2003); hep-ph/0212219.
\bigskip

\noindent [11] A. Melchiorri, in $Proceedings$ $of$ $the$ $I.A.P.$ $Conference$ ``$On$ $the$ $Nature$ $of$ $Dark$ $Energy$'' Paris, 2002, edited by P. Brax, J. Martin and J.P. Uzan. 

\bigskip

\noindent [12] B. Ratra and P. J. E.Peebles, Phys.Rev. {\bf D 37}, 3406(1988); C. Wetterich, Nucl. Phys. {\bf B 302}, 668 (1988); J. Frieman, C.T. Hill, A. Stebbins and I. Waga, Phys. Rev. Lett. {\bf 75}, 2077 (1995); P. G. Ferreira and M. Joyce, Phys.Rev. {\bf D 58}, 023503(1998); I. Zlatev, L. Wang and P. J. Steinhardt, Phys. Rev. Lett. {\bf 82}, 896 (1999); P. Brax and J. Martin,  Phys.Rev. {\bf D 61}, 103502(2000);
 L. A. Ure${\tilde n}$a-L${\acute o}$pez and T. Matos Phys.Rev. {\bf D 62}, 081302(R) (2000); T. Barrriro, E. J. Copeland and N.J. Nunes, Phys.Rev. {\bf D 61}, 127301 (2000); A. Albrecht and C. Skordis,  Phys. Rev. Lett. {\bf D 84}, 2076 (2000); V. B. Johri,  Phys.Rev. {\bf D 63}, 103504 (2001); J. P. Kneller and L. E. Strigari, astro-ph/0302167; F. Rossati, hep-ph/0302159; V. Sahni, M. Sami and T. Souradeep,  Phys. Rev. {\bf D 65}, 023518 (2002); M. Sami, N. Dadhich and Tetsuya Shiromizu, hep-th/0304187.

\bigskip

\noindent [13] C. Armendariz-Picon, T. Damour and V. Mukhanov, Phys. Lett. {\bf B 458}, 209 (1999); T. Chiba, T. Okabe and M. Yamaguchi, Phys. Rev. {\bf D 62}, 023511 (2000) 

\bigskip

\noindent [14] A. Kamenshchik, U. Moschella and V. Pasquir, Phys. Lett. B {\bf 511}, 265 (2001); N. Bilic, G.B. Tupper and R. Viollier, $ibid.$ {\bf 535}, 17 (2002).

\bigskip

\noindent [15]  L. P. Chimento, A. S. Jakubi, D. Pav${\acute o}$n and W. Zimdahl, Phys. Rev. D {\bf 67}, 083513 (2003); $ibid$ {\bf 67}, 087302 (2003);  W. Zimdahl, D. Pav${\acute o}$n and L. P. Chimento, Phys. Lett. B {\bf 521}, 133 (2001); W. Zimdahl and D. Pav${\acute o}$n, Gen. Ralativ. Gravit. {\bf 35}, 413 (2003); D. Tocchini- Valentini and L. Amendola, Phys. Rev. D {\bf 65}, 063508 (2002).

\bigskip

\noindent [16] E. Gunzig $et$ $al$, Phys. Rev. D {\bf 63}, 067301 (2001); V. Faraoni, Int. J. Mod. Phys. D {\bf 11}, 471 (2002);  L. P. Chimento, A. S. Jakubi, D. Pav${\acute o}$n and W. Zimdahl, Phys. Rev. D {\bf 69}, 083511 (2004).

\bigskip

\noindent [17] V.Faraoni, Phys. Rev.D{\bf 68},063508 (2003) ; R.A. Daly $et$ $al.$, astro-ph/0203113; R.A. Daly and E.J. Guerra, Astron. J. {\bf 124} (2002) 1831;  R.A. Daly, astro-ph/0212107; S. Hannestad and E. Mortsell, Phys. Rev.D{\bf 66} (2002) 063508; A. Melchiorri $et$ $al.$, Phys. Rev.D{\bf 68} (2003) 043509; P. Schuecker $et$ $al.$, astro-ph/0211480;   R.R. Caldwell, Phys. Lett. B {\bf 545} (2002) 23; R.R. Caldwell, M. Kamionkowski and N.N. Weinberg, Phys. Rev. Lett. {\bf 91} (2003) 071301;   H. Ziaeepour, astro-ph/0002400 ; astro-ph/0301640; P.H. Frampton and T. Takahashi,  Phys. Lett. B {\bf 557} (2003) 135;  P.H. Frampton, hep-th/0302007;  S.M. Carroll $et$ $al.$, Phys. Rev.D{\bf 68} (2003) 023509;  J.M.Cline $et$ $al.$, hep-ph/0311312;  U. Alam , $et$ $al.$, astro-ph/0311364 ; astro-ph/0403687; O. Bertolami, $et$ $al.$, astro-ph/0402387;  P. Singh, M. Sami $\&$ N. Dadhich, Phys. Rev.D{\bf 68} (2003) 023522; M. Sami $\&$ A. Toporesky, gr-qc/0312009;  B. McInnes, JHEP, {\bf 08} (2002) 029; hep-th/01120066;  V. Sahni $\&$ Yu.V.Shtanov, JCAP {\bf 0311} (2003) 014; astro-ph/0202346;  Pedro F. Gonz$\acute a$lez-D$\acute i$az, Phys. Rev.D{\bf 68}, 021303(R) (2003); J. D. Barrow, Class. Quan. Grav. {\bf 21}, L 79 - L82 (2004) [gr-qc/0403084]; S. Nojiri and S. D. Odintsov, hep-th/0303117; hep-th/0306212; E. Elizalde, S. Noriji and S. D. Odintsov,  Phys. Rev.D{\bf 70}, 043539 (2004); hep-th/0405034, S. Nojiri and S. D. Odintsov, Phys. Lett. B {\bf 595}, 1 (2004); hep-th/0405078; hep-th/0408170 ; V. K. Onemli and R. P. Woodard, Class. Quant. Grav. {\bf 19}, 4607 (2002) [gr-qc/0204065]; gr-qc/0406098; T. Brunier, V. K. Onemli and R. P. Woodard, gr-qc/0408080 ;  S. K. Srivastava, astro-ph/0407048.

\bigskip
\noindent [18]J. D. Barrow, Class. Quant. Grav., {\bf 21}, L79 (2004);  gr-qc/0403084; gr-qc/0409062.

\bigskip

\noindent [19] K. Lake , gr-qc/0407107.

\bigskip

\noindent [20]  S. Nojiri and S. D. Odintsov, hep-th/0303117.

\bigskip

\noindent [21] T. Padmanabhan, Phys. Rep. {\bf 380}, 235 (2003).

\bigskip

\noindent [22] V. Sahni, astro-ph/0403324.

\bigskip

\noindent [23] S. Perlmutter $et$ $al$., Astrophys. J., {\bf 517}, 565 (1999).

\bigskip

\noindent [24] A.G. Riess $et$ $al$., Astron. J., {\bf 116}, 1009 (1998).

\bigskip

\noindent [25] J. Tonry $et$ $al$, Astrophys. J. {\bf 594}, 483 (2003).

\bigskip

\noindent [26] D. N. Spergel $et$ $al$., astro-ph/0302209.

\bigskip

\noindent [27] L. Page $et$ $al$., astro-ph/0302220.

\bigskip

\noindent [28] A. B. Lahnas, N. E. Nanopoulos and D. V. Nanopoulos, Int. Jour. Mod. Phys. D, {\bf 12}, 1529 (2003).

\bigskip

\noindent [29] P. J. Steinhardt, Cosmological challenges for the 21st century, in : V. L. Fitch, D.R.Marlow (edts.), Critical Problems in Physics, Princeton Univ. Press, Princeton, NJ, 1997. 

\bigskip

\noindent [30] T. Padmanabhan,  Phys. Rev.D{\bf 66}, 021301(R) (2002).

\bigskip

\noindent [31]  L. P. Chimento, A. S. Jakubi and D. Pav${\acute o}$n , Phys. Rev. D {\bf 62}, 063508 (2000)

\bigskip

\noindent [32] R. Herrera, D. Pav${\acute o}$n and W. Zimdahl, astr0-ph/0404086.

\bigskip

\noindent [33] S. K. Srivastava, hep-th/0404170.

\bigskip

\noindent [34] D. F. Mota and C. van de Bruck, Astro. Astrophys., {\bf 421}, 71 (2004); astro-ph/0401504.

\end{document}